\documentclass[showpacs,preprintnumbers,amsmath,amssymb]{revtex4}
\usepackage{graphicx}
\usepackage{dcolumn}
\makeatletter
\input{epsf}
\begin{document}
\title{ Fractional Quantum Hall Edge Electrons,
Chiral Anomaly and Berry Phase}
\author{B. Basu}
 \email{banasri@isical.ac.in}

\author{P. Bandyopadhyay}
 \email{pratul@isical.ac.in}
\affiliation{Physics and Applied Mathematics Unit\\
 Indian Statistical Institute\\
 Kolkata-700108 }
\begin{abstract}
\begin{center}
Abstract
\end{center}
It is shown that the deviation of fractional quantum Hall edge
fluid from power law correlation functions with universal exponent
$\alpha=1/\nu$ as observed in recent experiment may be explained
when analyzed from the viewpoint of chiral anomaly and Berry
phase. It is observed that at the edge anomaly vanishes and this
induces a nonlocal effect in the construction of the electron
creation operator in terms of the edge boson fields. This
nonlocality is responsible for the deviation of the power law
exponent from $\alpha=1/\nu$ of the edge fluid. There are gapless
edge excitations described by chiral boson fields and the number
of branches of  chiral boson fields can be related to the
polarization states of electrons in the bulk. We have noted that
for fully polarized state at $\nu=\frac{1}{2m \pm 1}$ we will have
a single branch whereas for partially polarized state at
$\nu=\frac{n}{2mn \pm 1}$, with $n
>1$ and odd, we will have $n$ branches. However for unpolarized
state at $\nu=\frac{n}{2mn \pm 1}$ with $n$ even we will have two
branches.
\end{abstract}
\maketitle
\section{introduction}
The edge of a noninteracting electron system at integer filling
factor constitutes at low energy a one dimensional chiral electron
system that can travel only in one direction. The property of
fractional quantum Hall (FQH) edge could be described
\cite{wen1,wen2,wen3} by chiral Luttinger liquid and edge
excitations could be represented microscopically in terms of a
number of chiral boson fields. Early work did support this idea
and appeared to imply that the edge structure of quantum Hall
systems at the Laughlin states with $\nu=\frac{1}{2p+1}$, $p$
being a positive integer, is well described in terms of a chiral
Luttinger liquid characterized by a power law exponent
$\alpha=\frac{1}{\nu}$. Recently however this universal dependence
of $\alpha$ has been questioned both theoretically
\cite{gold,tsiper,man,wan} and experimentally\cite{hilke}. Besides
tunnelling density of states observations \cite{gray} at
hierarchical filing factors with $\nu=\frac{n}{2pn \pm 1}$ with
integer $n>1$ are found to be in apparent contradictions with the
predictions \cite{kane} of the chiral Luttinger liquid theory.

To explain this discrepancy Zulicke, Palacios and Macdonald
\cite{zu} have pointed out that it may not be a priori obvious
that electrons created in the low energy edge state Hilbert space
sector of incompressible fractional quantum Hall states satisfy
Fermi statistics. In the composite fermion framework \cite{man2}
it is suggested that a relevant perturbation caused by the weak
residual interaction between composite fermions may be responsible
for this discrepancy in the universal behaviour in the power law
exponent observed in experiments. Indeed, it has been pointed out
that the electron field operator may not have a simple
representation in the effective one dimensional theory and a
nonlocal effect may crop up in this one dimensional problem.

In this note we shall study the edge states of FQH incompressible
quantum Hall liquid from an analysis of the hierarchical quantum
Hall systems studied in the framework of chiral anomaly and Berry
phase \cite{pb1,pb2}. In sec. 2 we shall recapitulate certain
aspects of the studies in hierarchical states from the viewpoint
of Berry phase. In sec. 3 we shall study edge states of FQH
systems in this framework and in sec. 4 we shall discuss the edge
modes at various filling factors.

\section{Fractional Quantum Hall States: A Berry Phase Approach}

In some earlier papers \cite{pb1,pb2} we have analyzed the
sequence of quantum Hall states from the viewpoint of chiral
anomaly and Berry phase. To this end, we have taken quantum Hall
states on the two dimensional surface of a $3$D sphere with a
magnetic monopole of strength $\mu$ at the centre. In this
spherical geometry, we can analyze quantum Hall states in terms of
spinor wave functions and take advantage of the analysis in terms
of chiral anomaly which is associated with the Berry phase. In
this geometry the angular momentum relation is given by

\begin{equation}
{\bf J} ~ = ~ {\bf r} \times {\bf p} - \mu {\bf \hat{r}} ,
~~~~~~~~ \mu ~=~ 0 , ~\pm ~1 / 2 , ~\pm ~1 , ~\pm ~ 3 /
2,~.........
\end{equation}

From the description of spherical harmonics $Y^{m, \mu}_{\ell}$
with $\ell=1/2,~|m|=|\mu|=1/2$, we can construct a two-component
spinor $\theta = \left(
\begin{array}{c}
                          \displaystyle{u}\\
                          \displaystyle{v} \end{array} \right)$~~where
\begin{equation}
\begin{array}{lcl}
u ~=~ Y^{1/2 , 1/2}_{1/2} &=& \displaystyle{sin ~\frac{\theta}{2}
\exp
\left[ i (\phi - \chi) / 2 \right]}\\ \\
v ~=~ Y^{- 1/2 , 1/2}_{1/2} &=& \displaystyle{cos
~\frac{\theta}{2} \exp \left[- i (\phi + \chi) / 2 \right]}
\end{array}
\end{equation}
Here $\mu$ corresponds to the eigenvalue of the operator $i
\frac{\partial}{\partial \chi}$.

The $N$-particle wave function for the quantum Hall fluid state at
$\nu=\frac{1}{m}$ can be written as
\begin{equation}
{\psi^{(m)}}_{N} ~=~ \prod_{i < j} {(u_i v_j - u_j v_i)}^{m}
\end{equation}
$m$ being an odd integer. Here $u_i(v_j)$ corresponds to the
$i$-th ($j$-th) position of the particle in the system.

It is noted that ${\psi^{(m)}}_{N}$ is totally antisymmetric for
odd $m$ and symmetric for even $m$. We can identify  \cite{hal}
$m$ as $m=J_i+J_j$ for the N-particle system where $J_i$ is the
angular momentum of the $i$-th particle. It is evident from eqn.
(1) that with ${\bf r} \times {\bf p}=0$ and $\mu=\frac{1}{2}$ we
have
 $m=1$ which corresponds to  the complete
filling of the lowest Landau level.
 From the Dirac quantization condition $e
\mu =\frac{1}{2}$, we note that this state corresponds to $e=1$
describing the IQH state with $\nu=1$.

The next higher angular momentum state can be achieved either by
taking ${\bf r} \times {\bf p}=1$ and $\mid\mu \mid=\frac{1}{2}$
(which implies the higher Landau level) or by taking ${\bf r}
\times {\bf p}=0$ and $\mid{\mu_{eff}} \mid=\frac{3}{2}$ implying
the ground state for the Landau level. However, with
$\mid{\mu_{eff}} \mid=\frac{3}{2}$, we find the filling fraction
$\nu=\frac{1}{3}$ which follows from the condition $e \mu
=\frac{1}{2}$ for $\mu=\frac{3}{2}$. Generalizing this, we can
have $\nu=\frac{1}{5}$ with $\mid{\mu_{eff}} \mid=\frac{5}{2}$. In
this way we can explain all the FQH states with
$\nu=\frac{1}{2m+1}$ with $m$ an integer.

As $\mu$ here corresponds to the monopole strength, we can relate
this with the Berry phase. Indeed $\mu=\frac{1}{2}$ corresponds to
one flux quantum and the flux through the sphere when there is a
monopole of strength $\mu$ at the centre is $2\mu$ . The Berry
phase of a fermion of charge $q$ is given by $e^{i\phi_B}$ with
$\phi_B=2\pi q N$ where $N$ is the number of flux quanta enclosed
by the loop traversed by the particle. It may be mentioned that
for a quantum Hall particle the charge is given by $-\nu e$ when
$\nu$ is the filling factor.

If $\mu$ is an integer, we can have a relation of the form
\begin{equation}
{\bf J} ~=~ {\bf r} \times {\bf p} - \mu {\bf \hat{r}} ~=~ {\bf
r}^{~\prime} \times {\bf p}^{~\prime}
\end{equation}
which indicates that the Berry phase associated with $\mu$ may be
unitarily removed to the dynamical phase. Evidently, the average
magnetic field may be considered to be vanishing in these states.
 The attachment of
$2m$ vortices ($m$ an integer) to an electron effectively leads to
the removal of Berry phase to the dynamical phase. So, FQH states
with  $2 \mu_{eff} = 2m + 1$  can be viewed as if one vortex is
attached to the  electron. Now we note that for a higher Landau
level we can consider the Dirac quantization condition $e
\mu_{eff} = \frac{1}{2} n$, with $n$ being a vortex of strength $2
\ell + 1$. This can generate FQH states having the filling factor
of the form $\frac{n}{2 \mu_{eff}}$ where both $n$ and $2
\mu_{eff}$ are odd integers. Indeed, we can write the filling
factor as \cite{pb1,pb2}
\begin{equation}
\nu ~=~\frac{n}{2 \mu_{eff}} ~=~ \frac{n}{(2 \mu_{eff}~ \mp 1) \pm
1}  ~=~\frac{n}{2m' \pm 1}~=~ \frac{n}{2mn \pm 1} \label{nu1}
\end{equation}
where $2 \mu_{eff} \mp 1$ is an even integer given by $2m'=2mn$.
The particle-hole conjugate states can be generated with the
filling factors given by
\begin{equation} \nu ~=~ 1 - \frac{n}{2mn \pm 1}
~=~ \frac{n (2m - 1) \pm 1}{2mn \pm
1}=\frac{n^{\prime}}{2mn^{\prime} \pm 1} \label{e3}
\end{equation}
where $n(n')$ is an odd(even) integer,

Now, to study the polarization of various FQH states  \cite{bsp}
we observe that in the lowest Landau level we have the filling
factor
 $$\nu=\frac{1}{2 \mu_{eff}}=\frac{1}{2m+1}$$
As we have pointed out the attachment of $2m$ vortices to an
electron leads to the removal of Berry phase to the dynamical
phase and this effectively corresponds to the attachment of one
vortex (magnetic flux) to an electron, this electron will be a
polarized one.

However, in the higher Landau level this scenario will change.
Here we can write
\begin{equation}\label{nu1}
  \nu=\frac{n}{2\mu_{eff}}=\frac{1}{\frac{2\mu_{eff} \mp 1}{n}
   \pm \frac{1}{n}} ~~~{\rm with}~ n>1~ {\rm and}~ {\rm odd}
\end{equation}
where $2\mu_{eff} \mp $ is an even integer. We note that as even
number of flux units can be accommodated to the dynamical phase we
may consider this as $\frac{1}{n}$ flux unit is attached to an
electron implying that $n$ electrons will share one flux. This
suggests that electrons will not be fully polarized as one full
flux unit is not available to it. This will correspond to
partially polarized states with $$\nu= \frac{n}{2mn \pm 1}~~n>1
~{\rm and~odd} $$ which represents the states like $3/5, 3/7$ and
so on.

For the states with the filling factor $$\nu=\frac{n'}{2mn' \pm
1}, ~~~ n'~ {\rm an~ even~ integer}$$ we have observed that this
is achieved when we have particle-hole conjugate states given by
$$\nu=1-\frac{n}{2mn \pm 1}~~ {\rm with}~ n ~{\rm an~ odd~ integer}$$
A hole configuration is described by the complex conjugate of the
particle state where the spin polarization of the particle and
hole states will be opposite to each other. This will represent an
unpolarized state.

\section{Edge States of Fractional Quantum Hall Liquid}

It is noted that in the framework of spherical geometry, there is
no edge. However we can take the flat limit and in that case the
edge will be characterized by the limiting case $\mu \rightarrow 0
$. Indeed, when Hall fluid is taken to reside on the $2D$ surface
of a $3D$ sphere of large radius with a monopole of strength $\mu$
at the centre we can take the flat limit such that the edge of the
surface is characterized by the fact that the effect of the
monopole at the boundary should be in the vanishing limit. In
fact, beyond the boundary away from the bulk the space is
isotropic and the angular momentum is just given by ${\bf
L}=\bf{r} \times \bf p$ so that on the boundary the effect of
$\mu$ is in the vanishing limit. In the bulk we have non-vanishing
$\mu$ associated with chiral anomaly given by the relation
\cite{dp}
\begin{equation}\label{chiral}
  q=2\mu=-\frac{1}{2} \int \partial_\mu {J_\mu}^5 d^4x
\end{equation}
where ${J_\mu}^5$ is the axial vector current
$\bar{\psi}\gamma_\mu \gamma_5 \psi $. Here $q$ is an integer
known as the Pontryagin index. We note that for vanishing $\mu$ at
the boundary we should have a chiral current associated with
${J_\mu}^5$ having the opposite chirality at the boundary. That
is, at the edge we have a chiral current such that the chirality
is opposite to that in the bulk. This suggests that there is an
extra edge current arising out of the boundary effect and edge
velocities appear as external parameters that are not contained in
the bulk effective theory. This implies that we should have
anomaly cancellation at the edge \cite{pb3}.

It was suggested that the edge state could be described by the
chiral Luttinger liquid \cite{wen1,wen2,wen3}. To have an insight
into this approach to describe a one dimensional system of chiral
fermions, we note that the commutator for the density operator is
given by
\begin{equation}\label{wen}
[\rho_{-q'}, \rho_q]= \frac{qL}{2\pi} \delta_{qq'}
\end{equation}
where $L$ is the length and $q$ the wave vector. Now we define
\begin{equation}\label{ope}
 a_q=-i\sqrt{\frac{2\pi}{qL}}\rho_{-q},~~~~
 {a\dag}_q=
i\sqrt{\frac{2\pi}{qL}}\rho_{q}
\end{equation}
which satisfy
\begin{equation}\label{com}
 [a_{q'},{a^\dag}_q]=\delta_{qq'}.
\end{equation}
Now one defines the bosonic field
\begin{equation}\label{bos}
  \phi(x)=\sum_{q>0}\sqrt{\frac{2\pi}{qL}}(e^{-iqx}a_q+e^{iqx}{a\dag}_q)
  e^{-a |q| /2}
\end{equation}
where $a$ is a regularization cut-off which is set to zero at the
end. The electron field operator can be written as
\begin{equation}\label{field}
  \psi_e (x) \sim e^{-i \phi(x)}~~~~~~~~~~~~~~~~~~~~~~~~~~~~~~~
\end{equation}
For the FQH edge state with filling factor $\nu=1/m$, $m$ being an
odd integer, it can be taken
\begin{equation}\label{ope1}
[\rho_{-q'}, \rho_q]=\frac{1}{m}\frac{qL}{2\pi} \delta_{qq'}
\end{equation}
so that the operators $a$ and $a\dag$ take the form
\begin{equation}\label{ope2}
  a_q=-i\sqrt{m}\sqrt{\frac{2\pi}{qL}}\rho_{-q},~~~~{a\dag}_q=
  i\sqrt{m}\sqrt{\frac{2\pi}{qL}}\rho_{q}
\end{equation}
Wen postulated that we can reconstruct the bosonic field $\phi(x)$
with these new operators when the electronic field is given by
\begin{equation}\label{field1}
  \psi_e (x) \sim e^{-i \sqrt{m}\phi(x)}~~~~~~~~~~~~~~~~~~~~~~~~~~~~
\end{equation}
 For  odd $m$, we have consistency with the antisymmetric
property
\begin{equation}\label{commu}
  \{\psi_e(x),\psi_e(x')\}=0
\end{equation}
Noting that $\phi$ is a bosonic field with the propagator
\begin{equation}
\left< \phi(x,t) \phi(0) \right> =\nu~ ln(x-vt)
\end{equation}
the electron propagator becomes
\begin{equation}
  G(x,t)=\left< T\psi^{\dag}(x,t)\psi(0)\right>
        = \exp(\frac{1}{\nu^2}<\phi(x,t)\phi(0)>
         \sim  \frac{1}{(x-vt)^m}
\end{equation}
with $m=\frac{1}{\nu}$. Thus the electron propagator on the edge
of a FQH state acquires a nontrivial exponent $m=\frac{1}{\nu}$
that is not equal to 1. This type of electron state is called a
chiral Luttinger liquid though  the recent experiments do not
support this prediction of chiral Luttinger liquid.

We have pointed out that in the framework of spherical geometry in
the flat limit the edge state is characterized by the limiting
procedure $\mu \rightarrow 0$ where the chiral anomaly vanishes.
To consider this limiting procedure we take resort to a
renormalization group analysis involving the factor $\mu$
\cite{pp,bpd}. As is well known, in the angular momentum relation
(1) $\mu$ can take values $0,\pm 1/2,\pm 1,\pm 3/2...$. To have a
renormalization group analysis we take $\mu$ not to be a fixed
value but dependent on a parameter. Indeed, we can consider a
function $\mu (\lambda)$ which satisfies\\
1) $\mu$ is stationary at some fixed values of $\lambda=
\lambda^*$ of the RG flow {\it i.e.}
$\nabla \mu (\lambda^*) = 0$\\
2) At the fixed points $\mu(\lambda^*)$ is equal to the Berry
phase factor
$\mu^*$ of the theory\\
3) $\mu$ is decreasing along the infrared (IR) RG flows i,e.
$L\frac{\partial \mu}{\partial L} \le 0$ where $L$ is a length
scale. \\

We now consider  magnetic flux quanta passing through a domain $D$
characterizing a length scale $L$ and let a three dimensional
smearing density function $f(a)$ be a positive decreasing function
such that $a \frac{\partial f} {\partial a} \le 0$. We now write
the expression for the gauge potential
 \begin{equation}
  [A_{\mu}(x)]_D = \int_D d^3 a f(a) {\tilde A}_\mu (x,a)
  \end{equation}

So from the expression which relates the Berry phase factor $\mu$
with the chiral anomaly given by \cite{dp}
\begin{equation}
 2 \mu =-1/2 \int \partial_{\mu} {J_\mu}^5 d^4x=
-\frac{1}{16\pi^2} \int~~^\ast F_{\mu\nu} (x) F_{\mu\nu} (x) d^4 x
\end{equation}
where $F_{\mu\nu}(x)$ being the field strength and $^\ast
F_{\mu\nu} (x)$ the Hodge dual, we can write for the {\it flux
density}
\begin{equation}
\mu = {[\int \tilde{\mu} (x) d^4 x]}_D
  = -\frac{1}{32 \pi^2} \int
d^4 x \int_D d^3 a f(a) ^\ast {\tilde F}_{\mu\nu} (x, a) {\tilde
F}_{\mu\nu} (x,a)
 \end{equation}
Here ${[\tilde{\mu}(x)]}_D$ effectively gives the smearing of the
pole strength over the domain $D$. The $\mu$-function defined
above is a pure number but now explicitly depends on the length
scale $L$ characterizing the size of the domain. Now noting that a
global change of scale $L$ for the off-critical model amounts to a
change of the coupling constant $\lambda^i \to \lambda^i(L)$, the
renormalization group flux equations can be written as
\begin{equation}
L\frac{\partial}{\partial L} \lambda^i = - \beta^i
\end{equation}
which suggests that
\begin{equation}
 -\beta^i \frac{\partial \mu}{\partial \lambda^i} =
 L \frac{\partial \mu}{\partial L}
 \leq 0
\end{equation}
 It is noted that $\mu$ takes the usual discrete values of $0,
\pm \frac{1}{2}, \pm 1, \pm \frac{3}{2} ...$ at fixed points of
the RG flows where $\mu$ is stationary and represents the Berry
phase factor $\mu^\ast$ of the theory. In terms of energy scale,
this suggests that as energy increases (decreases) $\mu$ also
increases (decreases).

From eqn.(24) it is noted that there is a length scale $L$ which
increases as $\mu$ approaches zero. We can consider this length
scale of the order of magnetic length separated away from the bulk
over which the limiting procedure $\mu \rightarrow 0$ persists.
This will induce a nonlocal effect suggesting that the electron
field operator $\psi_e(x)$ is represented by a nonlocal operator
in the edge state. In view of this we may write
\begin{equation}\label{local}
  \psi_e(x)=\int dy~ f(|y-x|)e^{i\sqrt{m}\phi(y)}
\end{equation}
where $f(|y-x|)$ is peaked at $y=x$. This will maintain the
antisymmetric relation (\ref{commu}) but the equal time Green's
function is given by
\begin{equation}\label{green}
  \left< {\psi ^{\dag}}_e(x)\psi_e(x')\right> \sim \int dy~\int dy'
  \frac{f(|y-x|)f(|y'-x'|)}{(y-y')^m}~~~~~~~~~~~~~~~
\end{equation}
If the function $f(|y-x|)$ is taken to be of the form $f(|y-x|)
\sim |y-x|^{-\beta}$, then we can write
\begin{equation}\label{green1}
\left< {\psi ^{\dag}_e}(x)\psi_e(x')\right> \sim |x-x'|^{-
\alpha}~~~~~~~
\end{equation}
with $\alpha = m-2(1-\beta)$. We have certain constraints for the
factor $\beta$. Indeed, the normalizability of $f(x)$ requires
$\beta~>\frac{1}{2}$ and to have well defined integrals at
coincident points we must have $\beta<1$. Thus $\alpha$ lies
between $m$ and $m-1$. Indeed, Mandal and Jain \cite{man} arrived
identical conclusion where nonlocality has been conjectured. It
may be observed that for $\nu=1/3$, the result is found to be
consistent with experiments.

We may now mention that in the local limit the field operator
$\psi_e(x)$ will not obey Fermi statistics as has been suggested
in [4]. Indeed, in the local limit, we take sharply $\mu=0$ which
implies the vanishing of chiral anomaly at the edge state in a
sharp manner. Now when there is no anomaly, the relation (8)
suggests that the Pontryagin index vanishes and hence we cannot
define a conserved quantum number like fermion number \cite{gp}.
In view of this analysis, we may suggest that the variation from
the Luttinger liquid prediction of correlation with power law
exponent $\alpha =1/\nu$ is related to a nonlocal effect at the
edge state.

\section{Edge Excitation Modes of FQH States}

The transport in fractional quantum Hall systems occurs at the
edges of the incompressible quantum Hall regions where gapless
excitations are present. However, the number of edge modes at
various filling factors and their dependence on the parameters
characterizing the FQH states is still an open question. It is not
clear how the number and chirality of boson branches in the edge
excitation spectrum is related to the characteristic features
associated with the incompressible fluid in the bulk. It has been
pointed out that \cite{dbc, ahm} due to electrostatic and exchange
considerations the edge of a FQH fluid divides into strips of
compressible and incompressible fluids. We shall study these
features in the present framework of our analysis of edge states.

In sec 2 we have observed that for the FQH states with filling
factor $\nu=\frac{1}{2m+1}$, electrons are fully polarized,
whereas for $\nu=\frac{n}{2mn\pm 1}$ with $n>1$ and odd, electrons
in the bulk are partially polarized. In case $\nu=\frac{n}{2mn\pm
1}$ with $n$ even, the system is unpolarized. Now for a fully
polarized state we can consider that in $ the~ clean ~limit$ the
system corresponds to a ferromagnet. Under the influence of an
external magnetic field we can consider the Heisenberg anisotropic
Hamiltonian representing nearest neighbour interaction which in 1D
can be written as
\begin{equation}\label{hei}
  H=J \sum ( \sigma_i^x \sigma_{i+1}^x +\sigma_i^y \sigma_{i+1}^y
  +\Delta \sigma_i^z \sigma_{i+1}^z)
\end{equation}
with $J<0$. Here,  $\Delta$ is the anisotropy parameter which may
be related to the monopole strength $\mu$ through the relation
$\Delta=\frac{2\mu+1}{2}$ \cite{pp}. It is noted that for
$\mu=1/2$ we have the isotropic Hamiltonian which is $SU(2)$
invariant. Now in the edge state as we have observed that in the
limit  $\mu \rightarrow 0$ the system undergoes a phase transition
and the Hamiltonian in this limit is given by
\begin{equation}\label{heisen}
  H=J \sum ( \sigma_i^x \sigma_{i+1}^x +\sigma_i^y \sigma_{i+1}^y)
  +\frac{J}{2} \sum \sigma_i^z \sigma_{i+1}^z
\end{equation}
It is noted that the Ising part of the Hamiltonian corresponds to
the near neighbour repulsion caused by free fermions. The (XY)
part of the Hamiltonian corresponds to a bosonic system which
dominates over the Ising part. When we translate this feature in
the edge fluid, we note that the bosonic part will give rise to a
strip of compressible fluid whereas the Ising part will contribute
to a strip of incompressible fluid. Thus in the edge state we will
have strips of compressible and incompressible fluid unlike in the
bulk where we have only incompressible fluid.

The Hamiltonian (\ref{heisen}) breaks the spin symmetry $SU(2)$
and this will give rise to chiral boson fields as the excitation
spectrum in the edge of the Hall fluid. The corresponding boson
fields will be chiral because these can move only in one
direction. Indeed at the filling factor $\nu=\frac{1}{2m \pm 1}$
we will have a single branch of edge excitation which corresponds
to the edge-magnetoplasmon (charged) mode. But the state with
filling factor $\nu=\frac{n}{2mn \pm 1}$ with $n>1$ and odd
corresponding to partially polarized state implies that the number
density of up spin and down spin electrons are unequal. We can
split this system in the bulk {\it in ~the~ clean~limit} in  $n$
branches of ferromagnetic systems. In the quantum Hall fluid the
edge will then give rise to $n$ branches of chiral boson fields as
excitations such that $n$ boson modes will share the same
chirality. That is these $n$ branches of edge excitations will
propagate along the same direction. In the same way we can suggest
that for unpolarized states where we have equal number of up and
down spin density and correspond to the filling factor
$\nu=\frac{n}{2m \pm 1}$ with $n$ an even integer, we will have
two branches of edge excitations sharing the same chirality. Thus
for FQH states with filling factors of the form $\nu=\frac{2}{4m
\pm 1}$, $\nu=\frac{4}{8m \pm 1}$..... and so on we will have two
branches of the edge excitations that share the same chirality.

We have pointed out that the edge state is governed by the
limiting procedure $\mu \rightarrow 0$ which is a nonlocal effect.
This non-locality is responsible for the observed deviation from
the chiral Luttinger liquid prediction of correlation having power
law exponent $\alpha =1/\nu$ of the edge fluid. It has been
suggested \cite{murthy} that the softening of the background
confining potential may lead to such discrepancy and this may
cause edge reconstruction leading to the deviation of the electron
density near the edge from the background charge profile. We may
suggest that the non-local effect at the edge may be related to
this softening of the confining potential and deviation of the
electron density profile which may persist in the region of a
magnetic length from the edge. However, numerical studies suggest
that quantum Hall edges at higher filling factors such that
$\nu=2/5,3/7$ are robust against reconstruction \cite{murthy}.
Indeed, these filling factors correspond to higher Landau levels
which necessitate more energy for electrons to occupy these
states. Now from the renormalization group equation (24), we note
that as the energy increases, $\mu$ also increases and so for a
length scale where $\mu$ decreases to the limiting value $\mu
\rightarrow 0$, the enhancement in energy scale will suppress this
effect implying that the non-local effect due to the limiting
procedure will be minimized. Indeed we will have only residual
non-local effect as we consider higher and higher Landau levels.
This will resist the edge reconstruction procedure and quantum
Hall liquid at the corresponding filling factor will be more
robust against the reconstruction as observed in numerical
studies.

\section{Discussion}

We have considered here a renormalization group analysis involving
the Berry phase factor $\mu$ to study the edge state of a
fractional quantum Hall fluid. It is observed that there is a
nonlocal effect in the construction of the electron creation
operator in terms of the edge boson fields. This non-locality may
be taken to be responsible for the deviation from chiral Luttinger
liquid power law correlation functions with universal exponent
$\alpha$ as observed in recent experiments. There are gapless edge
excitations and we have different branches of chiral bosonic
fields as excitations for various filling factors. We have noted
that for fully polarized state with $\nu=\frac{1}{2m \pm 1}$ we
will have a single branch whereas for partially  polarized state
at $\nu=\frac{n}{2mn \pm 1}$ with $n>1$ and odd  we will have $n$
 branches. However for unpolarized state
at $\nu=\frac{n}{2mn \pm 1}$ with $n$ even  we will have two
 branches.

 It may be mentioned here that recently some novel generation of filling
 factors for FQH fluid has been observed \cite{pan} which do not
 satisfy the primary Jain scheme. An analysis \cite{bp}
 of these states from
 a Berry phase approach suggests that the observed filling factors
 like $\nu= 4/11, 5/13,6/17, 4/13$ and $5/17$ correspond to
 the lowest Landau level whereas the state like $\nu=7/11$ correspond
 to particle-hole conjugate state. This implies that the former
 states will be fully polarized whereas the state with $\nu=7/11$
 will be unpolarized. Indeed this is found to be consistent with
 experiments for the state $\nu=4/11$ and $7/11$ \cite{pan}. So from
 our above analysis we may infer that the edge states for
 $\nu=4/11, 5/13, 6/17, 4/13$ and $5/17 $ will haves a single branch
 of bosonic excitation whereas for the state $\nu=7/11$ the edge
 excitations will have two branches of bosonic fields sharing the
 same chirality.

\end{document}